\documentclass{aa}

\usepackage{times}
\usepackage{graphics}
\usepackage{psfig}

\def\ltsima{$\; \buildrel < \over \sim \;$} 
\def\simlt{\lower.5ex\hbox{\ltsima}}
\begin{document}

\title{A multi-epoch spectrophotometric atlas of symbiotic 
       stars\thanks{Based on observations collected with the telescopes
       of the European Southern Observatory (ESO, Chile) and of 
       the Padova \& Asiago Astronomical Observatories 
       (Italy)}$^,$\thanks{Figures 4--256 are only 
       available in electronic form ($a$) at the CDS via anonymous 
       ftp to cdsarc.u-strasbg.fr (130.79.128.5) or via http://cdsweb.u-strasbg.fr/Abstract.html, 
       and ($b$) from the personal home page http://ulisse.pd.astro.it/symbio-atlas/ }}

\author{Ulisse Munari\inst{1,2} and Toma\v z Zwitter\inst{3}}

\offprints{Ulisse Munari}

\institute{Osservatorio Astronomico di Padova, Sede di Asiago, I-36032 Asiago
           (VI), Italy \ \ [{\tt munari@pd.astro.it}] 
\and       CISAS - Centro Inter-Dipartimentale per Studi ed Attivit\`a 
           Spaziali, Univ. di Padova,
	   Italy 
\and       University of Ljubljana, Department of Physics, 
           Jadranska 19, 1000 Ljubljana, Slovenia \ \ 
           [{\tt tomaz.zwitter@uni-lj.si}]}

\date{Received 9 October 20001 / Accepted 22 November 2001}

\markboth{Munari and Zwitter: A multi-epoch spectrophotometric atlas of
symbiotic stars}{Munari and Zwitter: A multi-epoch spectrophotometric 
atlas of symbiotic stars} 

\abstract{
A multi-epoch, absolute-fluxed spectral atlas extending from about 3200 to
9000~\AA\ is presented for 130 symbiotic stars, including members of the
LMC, SMC and Draco dwarf galaxies. The fluxes are accurate to better than
5\% as shown by comparison with Tycho and ground-based photometric data. The
spectra of 40 reference objects (MKK cool giant standards, Mira and Carbon
stars, planetary nebulae, white dwarfs, hot sub-dwarfs, Wolf-Rayet stars,
classical novae, VV~Cep and Herbig Ae/Be objects) are provided to assist
the interpretation of symbiotic star spectra. Astrometric positions and
counterparts in astrometric catalogues are derived for all program symbiotic 
stars. The spectra are available in electronic form from the authors.
}

\maketitle

\section{Introduction} 

Allen's (1984) atlas offered the first comprehensive spectroscopic view of
symbiotic stars. It included spectra covering the 3400-7500 \AA\ range for
114 objects (validated and possible, corresponding to 72\% of the total
number known at the time), which however were not fluxed, of low resolution
and low dynamic range. Nevertheless, Allen's catalogue has been extensively
used in all studies of symbiotic stars, even in those concerning single
objects, because its single-epoch spectra could be compared with later
observations in order to study the marked spectral variability of these
binaries.

Other compilations of symbiotic star optical spectra are available. The
larger ones since Allen's atlas are Blair et al. (1983; 16 objects, absolute
fluxes), Ipatov \& Yudin (1986; 14, absolute), Kenyon \& Fernandez-Castro
(1987; 11, absolute), Acker et al. (1988; 10, relative) M\'{u}rset et al.
(1996; 12, absolute), Gutierrez-Moreno et al. (1999; 14, absolute) and
Medina Tanco \& Steiner (1995; 45, relative). Meier et al. (1994) assembled
an atlas of ultraviolet IUE spectra for 32 symbiotic stars, Schulte-Ladbeck
(1988) of 16 objects in the near-IR and Schild et al. (1992) of 8 objects in
the IR. Van Winckel et al. (1993) surveyed emission line profiles for 59
objects and Ivison et al. (1994) for 35. Pereira et al. (1999) presented
Bowen-fluorescence dominated blue spectra for 8 symbiotic stars and Schmid
and Schild (1994) surveyed Raman-scatter dominated red spectra for 15
objects.

Here we present the largest optical spectroscopic atlas of symbiotic stars
since Allen's one. Compared to Allen's atlas, our spectra cover a
comparable number of objects (130 in all, 75 observed from ESO, 40 from
Asiago, and 15 from both places) but are better resolved , extend over a
wider wavelength range, offer a higher dynamical range, are absolutely
fluxed, are multi-epoch (half of the targets re-observed one or more times
over a three year period), and are available in electronic form (from 
the authors upon request).

The spectra presented in this atlas are planned to become the input data 
for future follow-up studies:\\
$-$ given the high accuracy of the absolute fluxes and the wide wavelength
range covered, optical magnitudes will be derived from the spectra and 
combined into a unique multi-epoch UBV(RI)$_{\rm C}$ photometric catalogue 
with the results of CCD photometry of 60 symbiotic stars obtained by Henden 
and Munari (2000, 2001a, 2001b) while calibrating their comparison 
sequences;\\
$-$ the rich emission line spectrum of symbiotic stars will be studied 
in a global approach;\\
$-$ properties of the cool giants will be derived from the absorption 
spectrum in the red region.

\section{The data} 

The list of program stars and the journal of observations is given in
Table~1. The number of symbiotic stars (validated and 

\begin{figure*}
\resizebox{18cm}{!}{\includegraphics{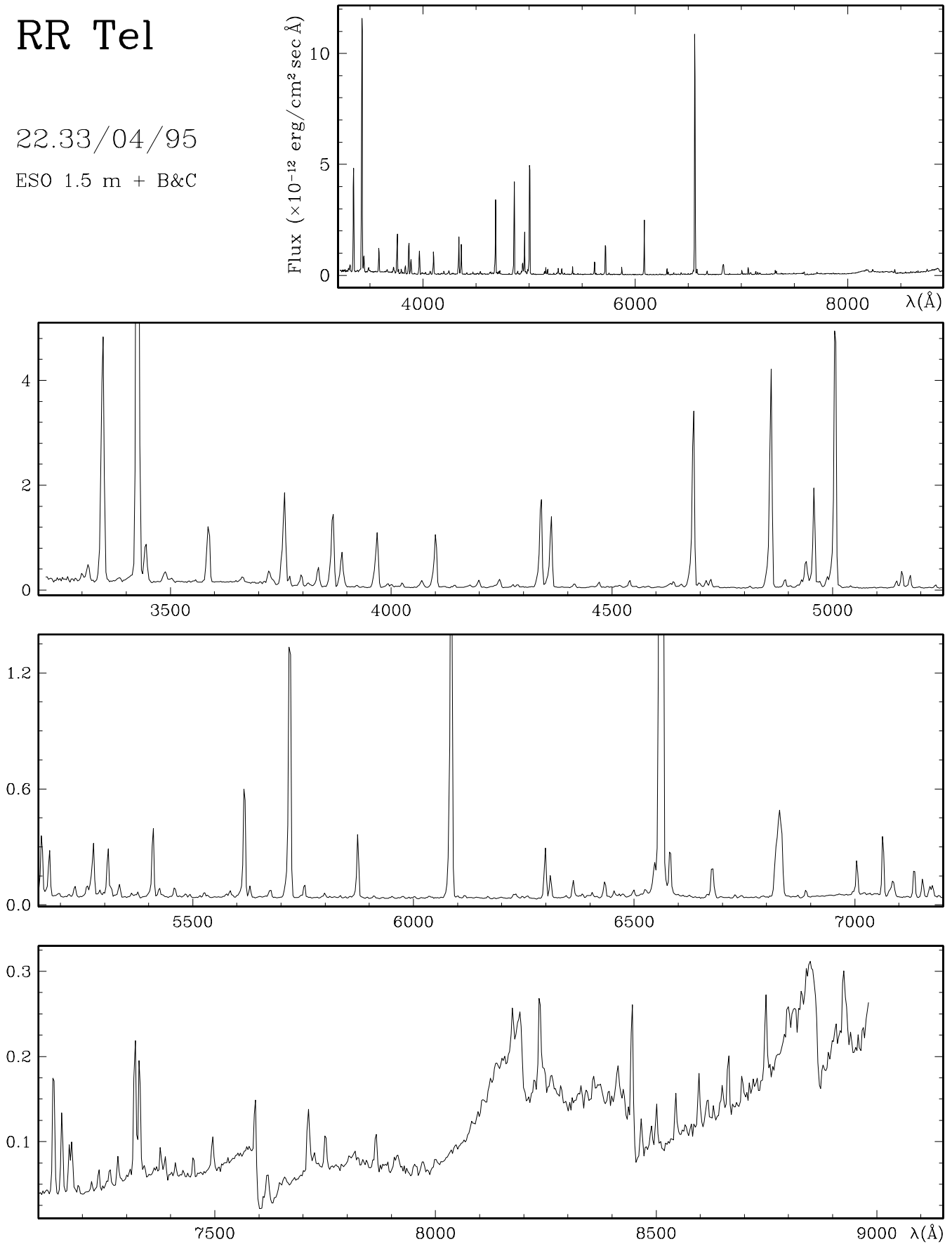}}
\caption{The spectrum of the symbiotic star RR~Tel. This is an example for 
the Figures 4-256 available electronically only.}
\end{figure*}

\clearpage

\begin{table*}[H!]
\caption[]{List of program stars and the journal of observations, 
           separately for ESO and Asiago observations.}
\centerline{\psfig{file=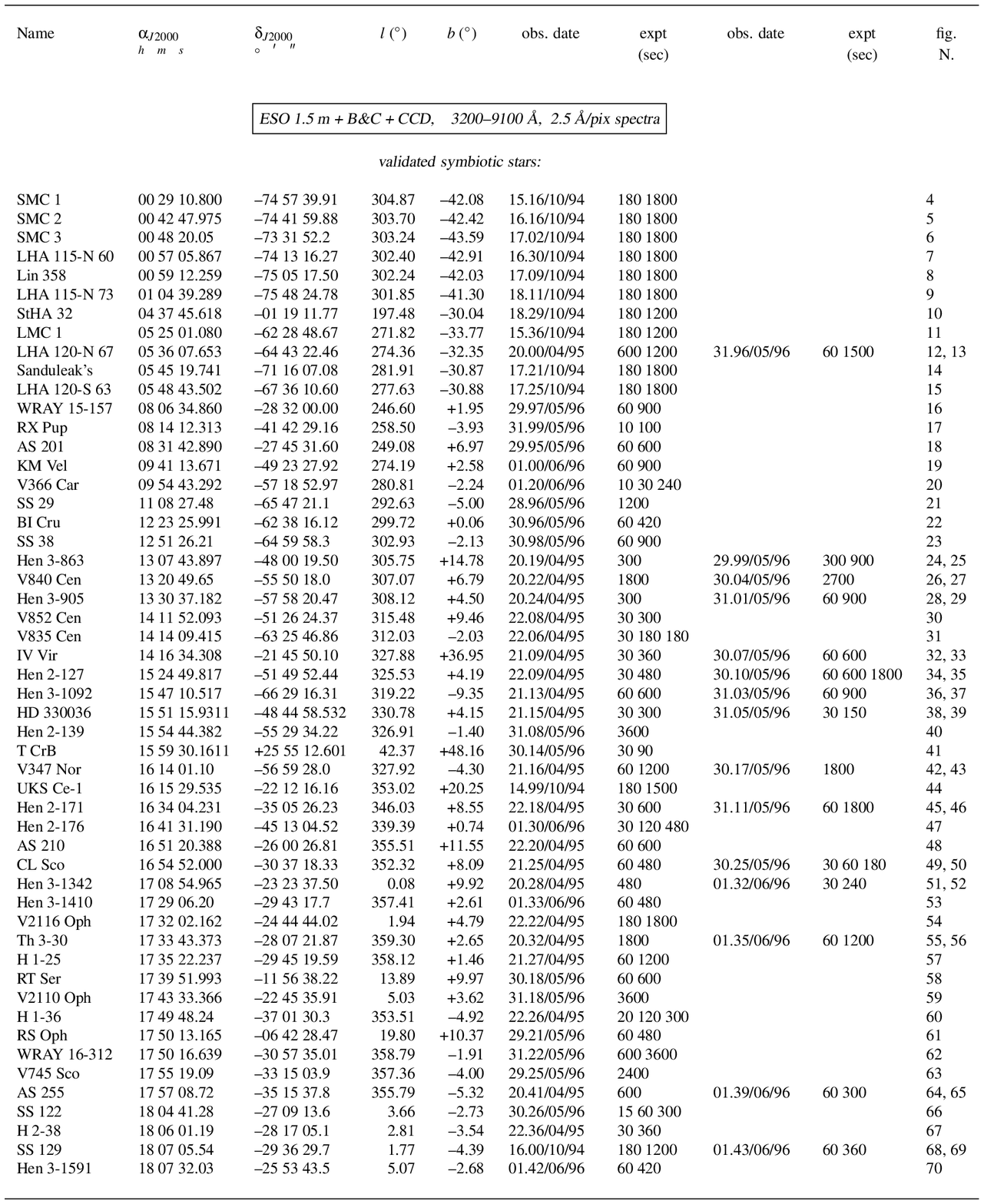,width=18cm}}
\end{table*}
\setcounter{table}{0}
\begin{table*}[H!]
\caption[]{({\sl continues})}
\centerline{\psfig{file=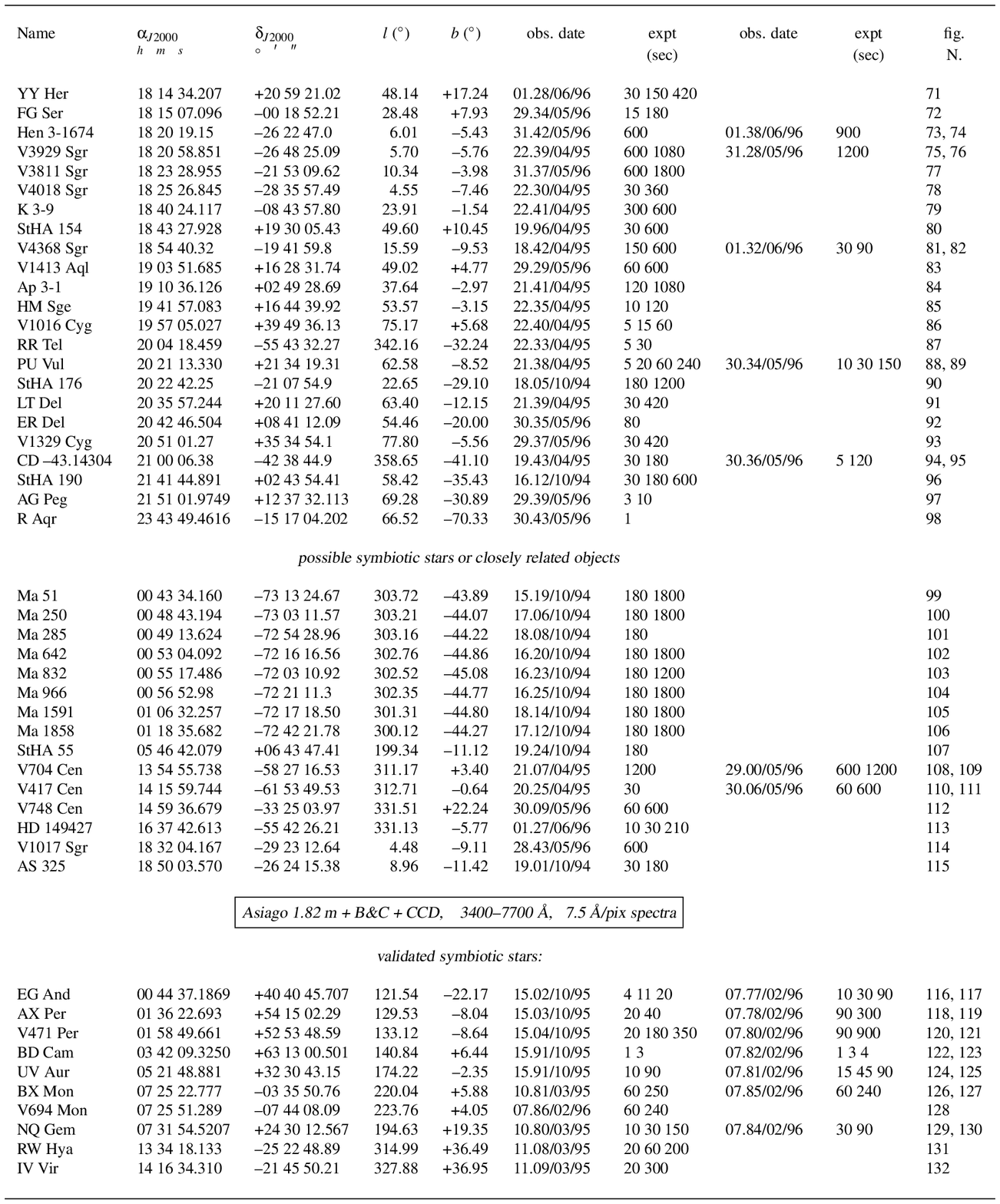,width=18cm}}
\end{table*}
\setcounter{table}{0}
\begin{table*}[H!]
\caption[]{({\sl continues})}
\centerline{\psfig{file=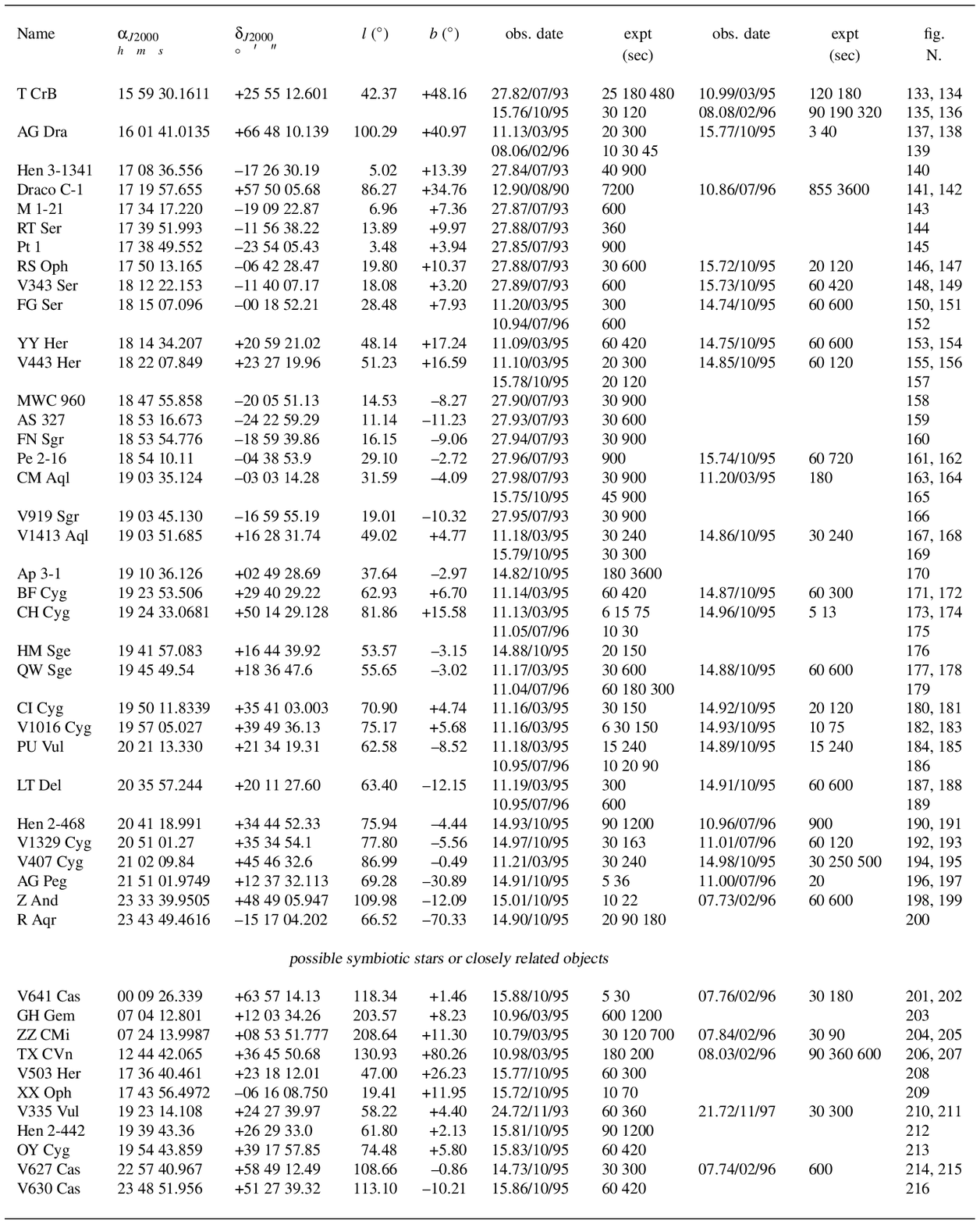,width=18cm}}
\end{table*}

\clearpage

\noindent
probable) observed for this atlas is 130, with a total of 213 spectra
presented in Figures~4--216. Spectra of 40 reference stars are given in
Figures~217--256, with Table~2 listing their relevant properties.  Figure~1
gives an example of Figures~4--256 available electronically only. Of the 90
symbiotic stars observed from ESO, 22 have been re-observed about one year
later. Fifty-five are instead the symbiotics observed from Asiago, 21 of
which were re-observed once, 9 twice and 1 three times over a two and a half
year period. For some systems our observations document the spectral changes
induced by an outburst state, as for CL~Sco (during the 1995 quiescence in
Fig.~49, and at 1996 outburst in Fig.~50), or their interplay with the
pulsation activity of the cool giant, as for the carbon symbiotic Mira
UV~Aur (at Mira's minimum in Fig.~124, and at Mira's maximum in Fig.~125).

The program stars have been selected among those listed in Allen's (1984)
catalogue and those discovered later (and summarized in the new catalogue of
symbiotic stars by Belczynski et al. 2000), including objects from the
Meyssonnier \& Azzopardi (1993) survey of SMC.

The object names (as given in Table~1 and the first column of Table~3) are
sometimes different from those used in the literature, because either (i) a
variable star name has been assigned since (for example V1413~Aql for
AS~338), or (ii) there have been re-organizations in the name coding within
SIMBAD (for example, He 2-468 was replaced by Hen 2-468, or Hen 1591 is now
Hen 3-1591). A list of correspondences is given in Table~3, where column~2
contains Allen (1984) names for the systems renamed subsequently.

The spectrophotometric observations have been performed with B\&C + CCD
spectrographs at ESO and Asiago. All spectra have been reduced with the IRAF
software package. Standard procedures (including bias and flat field
correction, wavelength and flux calibration) have been applied. Cosmic rays
on the stellar tracing were cleaned manually on the extracted spectrum after
inspection of the original two dimensional frames. For the vast majority of
the program stars we have obtained more than one spectrum per night (with
the aim of appropriately exposing the continuum as well as avoiding
saturation of the strongest emission lines), and their inter-comparison
greatly assisted during the manual cleaning of cosmic-rays.

\subsection{Astrometry}

Historically inaccurate or erroneous coordinates of several symbiotic stars
are still present in the current literature. Moreover, correspondence
between symbiotic stars and entries in astrometric catalogues has not been
systematically explored yet. The latter is an important task because, once
the correspondence is established, coordinates of symbiotic stars will be
automatically improved every time the catalogues are re-calibrated (like
USNO-A1 that has been recalibrated into USNO-A2 when the Hipparcos/Tycho
reference stars have become available) or cross-referenced toward newer and
higher precision catalogues.

Correspondence with sources in astrometric catalogues is established in
Table~3 (last column) for all the program stars.
Help has been provided by the {\sl Aladin} graphical interface at CDS
(http://aladin.u-strasbg.fr/aladin.gml) that allows overplotting of
astrometric catalogues over the digitized DSS-I and DSS-II plates. The
astrometric identification has been searched for and taken from the
following catalogues, in order of preference: Hipparcos, Tycho-2, USNO-A2.0,
GSC-1, 2MASS. Only 6 objects lack any astrometric identification, while 14
and 24 objects are included in the Hipparcos and Tycho-2 catalogues,
respectively.

Coordinates of program stars in Table~1 have not been adapted or
re-processed from the existing literature, but re-compiled from scratch.
Hender and Munari (2000, 2001a) give astrometrically measured accurate
positions for 40 of our program stars (linked to the USNO-A2.0 reference
system). For the remaining targets the positions as given in the appropriate
astrometric catalogues (following identifications in Table~3) have been
adopted. For the 6 objects without an entry in the surveyed astrometric
catalogues, coordinates have been measured by us on the digitized DSS-II
plates (or, if not available, on the DSS-I ones). All coordinates are on
J2000.0 equinox, but epochs are those of the corresponding catalogues (this
should not be a major concern given the minimal proper motions typical for
most symbiotic stars).

\subsection{ESO observations}

The ESO observations have been performed with the B\&C + CCD spectrograph at
the 1.5 m telescope under photometric conditions (sparse data from non
optimal nights have not been used in the present atlas). We used a 400 l/mm
grating (\# 25) and a 2 arcsec slit, always aligned along the parallactic
angle for pointings at zenital distances larger than 35$^\circ$. The CCD was
a 2048$\times$2048, 15 $\mu$m size, thinned and back-illuminated to enhance
UV sensitivity. The dispersion was $\sim$2.5 \AA/pix, with a
FWHM(PSF)$\sim$2 pixels. The covered spectral range changes somewhat from
one observing run to another, with the extremes of 3200--8900 and 3400--9100
\AA.  The seeing during the observing nights was always smaller than the
slit width (as can be derived from the simultaneous absolute measurements
with the La Silla Meteo Monitor and from the measured FWHM perpendicular to
the dispersion on the recorded stellar spectra). For a non--marginal part of
the time the seeing was better than 1 arcsec. Several spectrophotometric
standards were observed each night in order to achieve absolute flux
calibration and monitor stable transparency conditions.

\begin{figure*}
\resizebox{18cm}{!}{\includegraphics{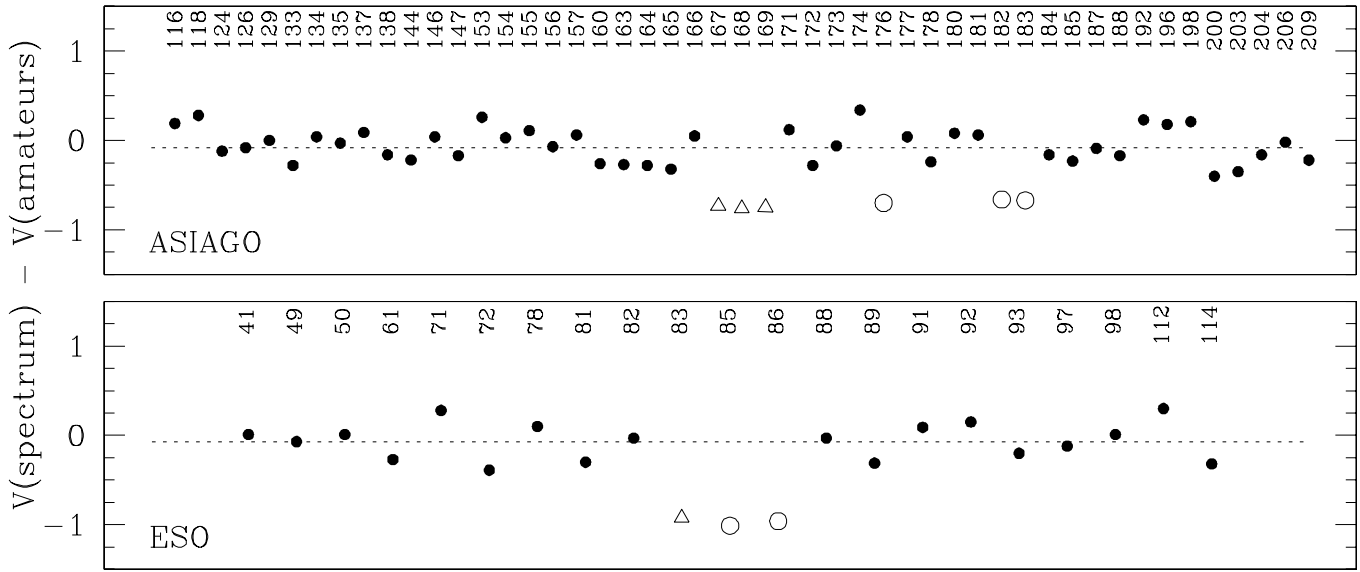}}
\caption{Comparison of $V$ magnitudes of symbiotic stars as estimated by
amateur astronomers (VSNET, VSOLJ, AFOEV databases) and as derived from 
flux calibrated spectra in this atlas. To identify the given symbiotic star 
and its particular spectrum, the upper row in each panel lists the 
corresponding Figure number (cf. Table~1). See text for details (sect. 2.6).}
\end{figure*}

\begin{figure*}
\resizebox{18cm}{!}{\includegraphics{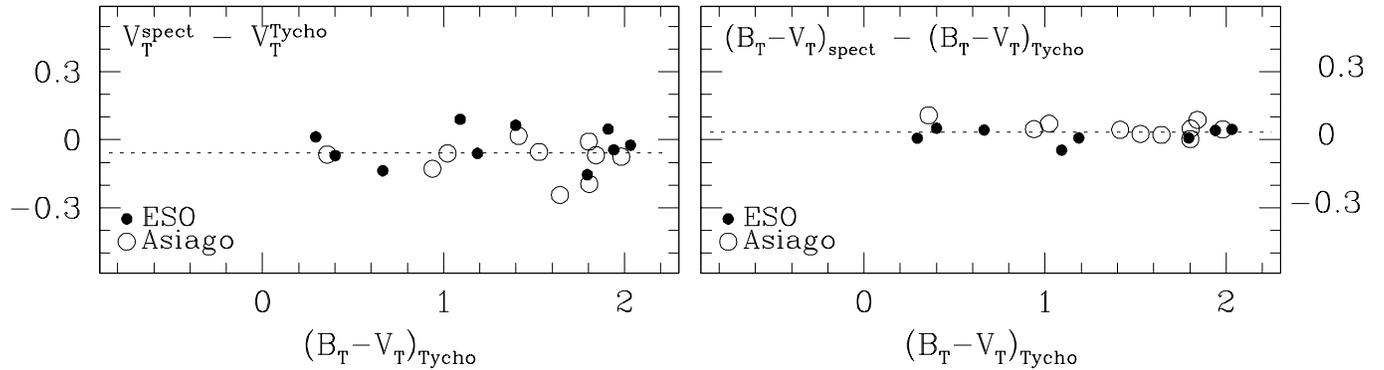}}
\caption{Comparison of $V_T$ magnitudes and ($B_T - V_T$) colors of template
stars in Table~2 as measured by Hipparcos/Tycho and as derived from spectra 
in this atlas. See text for details (sect. 2.6).}
\end{figure*}

\subsection{Asiago observations}

The Asiago observations were secured with the Boller \& Chivens spectrograph
attached to the 1.82 m telescope operated by Osservatorio Astronomico di
Padova atop of Mount Ekar (Asiago, Italy), during nights of suitable
photometric conditions. The detector was a Thompson TH7882 UV-coated CCD,
580x388 pixels of 23$\mu$m size. We used a 150 ln/mm grating giving a
dispersion of 7.5 \AA/pixel, generally covering the wavelengths 3350-7550
\AA\ (exact limits variable according to the observing runs). The slit width
was $\sim$2.0 arcsec, giving a FHWM(PSF)$\sim$2 pixels. When the object's
zenith distance exceeded 45$^\circ$, the slit was aligned along the
parallactic angle. Each night at least four spectrophotometric standard
stars were observed more than once at different airmasses for flux
calibration.

\subsection{Reference objects}

A number of spectra of reference objects have been secured with the same
instrumentation during symbiotic star observing runs. The reference objects
include MKK cool giant standards, Miras, planetary nebulae, Wolf Rayet
stars, white dwarfs, hot sub-dwarf, classical novae, and are intended to
assist inspection and interpretation of symbiotic star spectra. They are
listed in Table~2 and their spectra are presented in Figures~217--256.

\subsection{Flux accuracy}

It appears appropriate to quantify the accuracy of absolute fluxes in this
atlas. For symbiotic stars no optical photometry was carried out
simultaneously with the spectroscopic observations. However, a number of
program symbiotic stars are regularly observed by several organizations of
amateur astronomers. We have consulted the on-line databases of VSNET, VSOLJ
and AFOEV and found useful data to support comparison with 70 of the spectra
presented in this atlas (49 Asiago spectra and 21 ESO spectra, reflecting
the predominance of amateurs in the northern hemisphere and higher
brightness of average symbiotic stars known in the northern hemisphere).
When possible the amateurs' lightcurves have been interpolated/extrapolated
to derive the visual magnitude for the date of the given spectral
observation. In a few cases an estimate was found in the amateurs' archives
for the exact date of the spectral observations. Such estimates have been
adopted unless an eye inspection of the whole light-curve rendered them
unreliable.

\begin{table*}
\caption[]{Spectra of template objects}
\centerline{\psfig{file=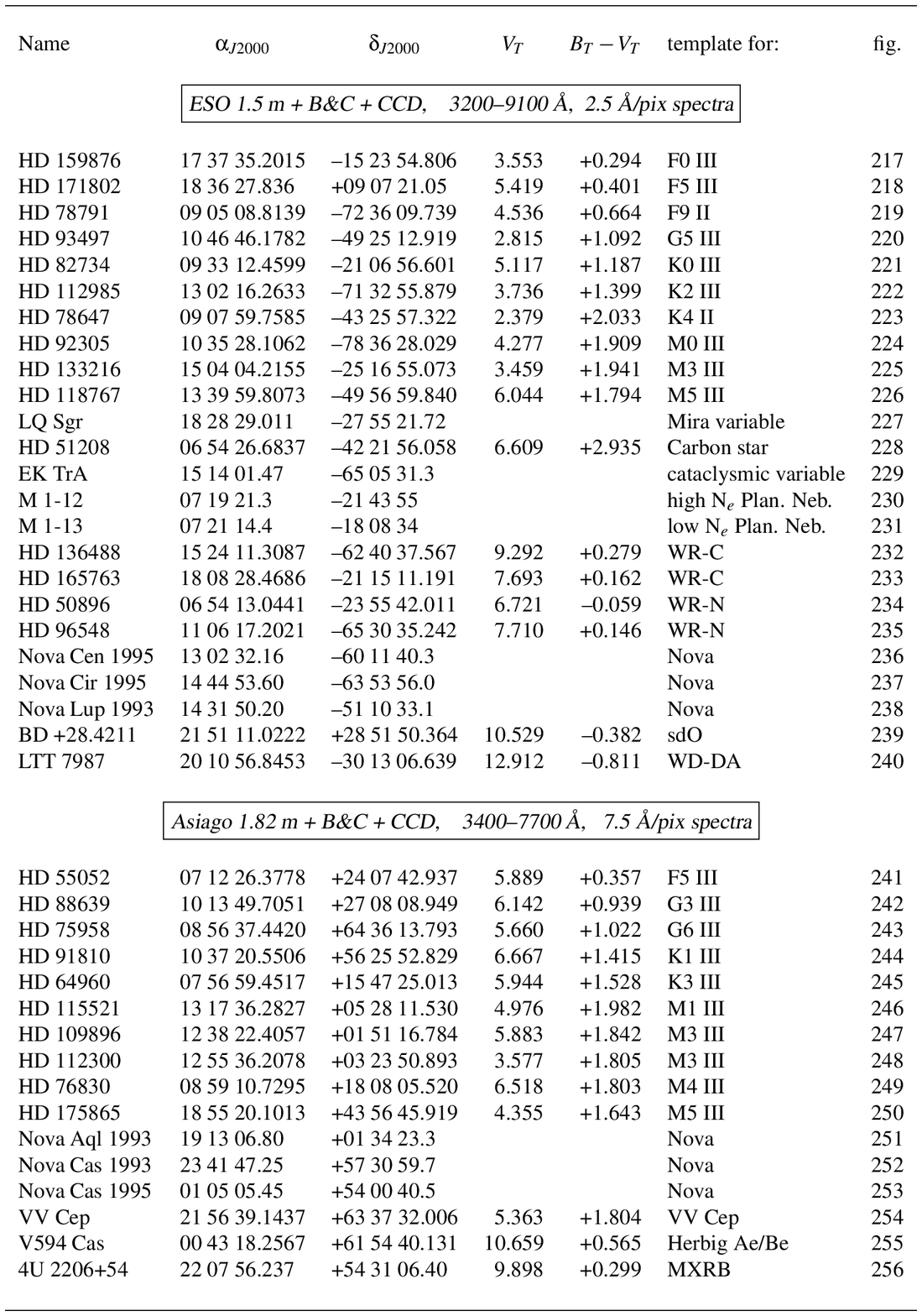,width=16cm}}
\end{table*}

\begin{table*}[h!]
\caption[]{Object names and astrometric reference}
\centerline{\psfig{file=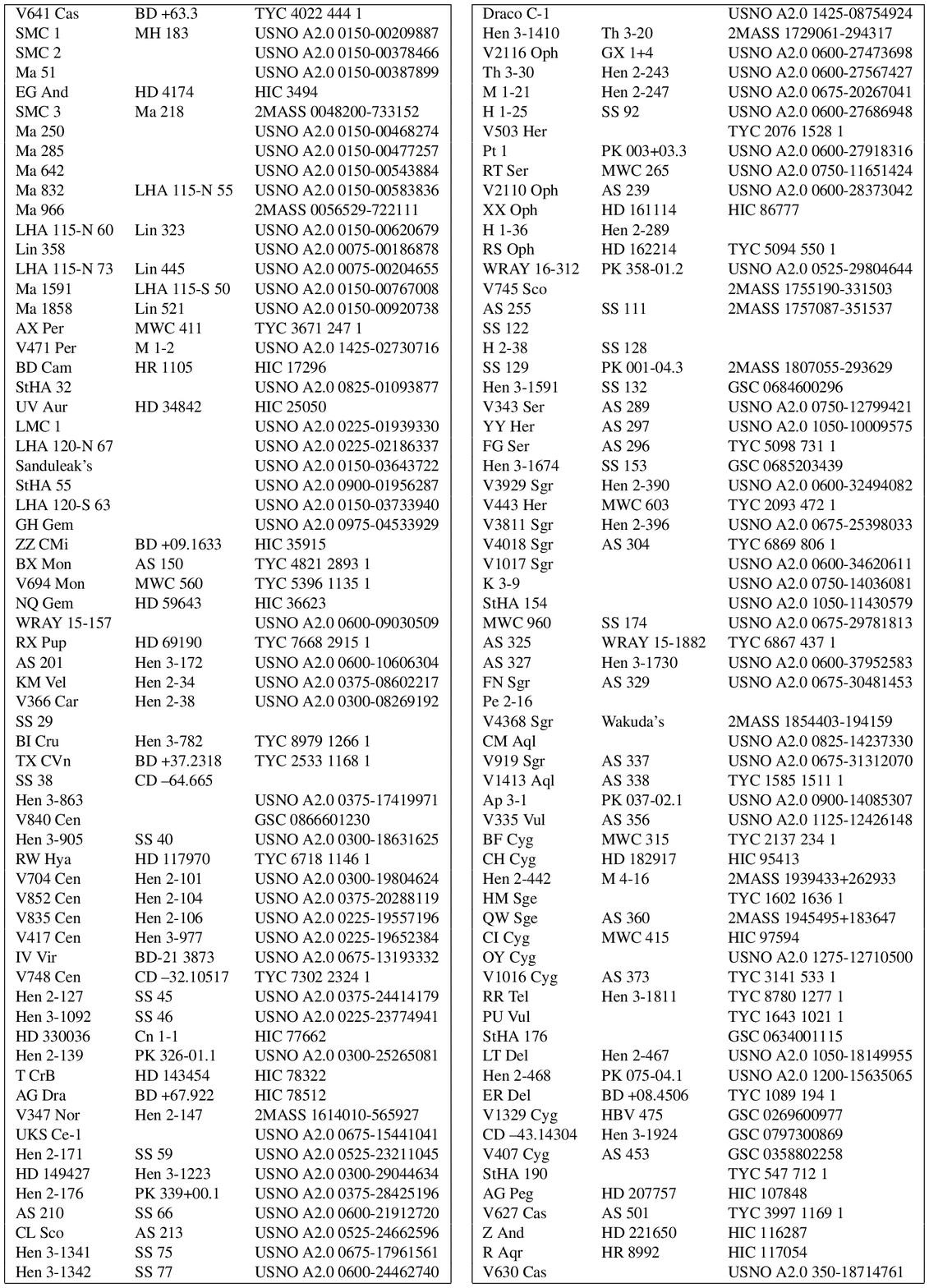,width=16cm}}
\end{table*}

\clearpage

The $V$ magnitude was derived from our spectra by convolving them with the
$V$ band-pass profile (taken from A\v{z}usienis \& Strai\v{z}ys 1969).  The
comparison between the spectroscopic and amateurs' $V$ magnitudes is
presented in Figure~2 separately for the Asiago and ESO data.
The comparison gives for the ESO spectra:
\begin{equation}
V_{spectrum}\ -\ V_{amateurs}\ =\ -0.06;\ \ \ \ \sigma\ =\ 0.21
\end{equation}
\noindent
and similarly for the Asiago spectra:
\begin{equation}
V_{spectrum}\ -\ V_{amateurs}\ =\ -0.05;\ \ \ \ \sigma\ =\ 0.19
\end{equation}
\noindent
The offset is small and can be easily accounted for by a number of
possibilities, the first and the most obvious one being the difference
between the eye response and the $V$ band profile.

ESO and Asiago observations appear equally accurate, as seen in Figures~2 and 3, 
with an average 0.20 mag scatter that deserves some
comment. Several mechanisms contribute to it, including [$a$] errors in
the comparison sequences used by amateurs ($\sigma_{cs}$), [$b$] errors in
estimating the magnitude by the amateurs and uncertainties in
interpolating/extrapolating their lightcurve to the desired date
($\sigma_{est}$), [$c$] day-by-day random variability of symbiotic stars
($\sigma_{var}$), and obviously [$d$] the errors in fluxing our spectra
($\sigma_{flux}$). Because these errors are unrelated and follow normal
distributions we write
\begin{equation}
\sigma^2\ =\ (0.20)^2\ =\ \sigma_{cs}^2\ +\ \sigma_{est}^2\ +\ \sigma_{var}^2\ 
           +\ \sigma_{flux}^2\
\end{equation}

Based on common experience it may be assumed that $\sigma_{cs} \geq$0.05,
$\sigma_{est} \geq$0.15, and $\sigma_{var} \sim$0.1 ($\sigma_{cs}=$0.05 is
obviously an underestimate, at least because magnitudes of comparison stars
on amateur finding charts are given to one decimal figure only). So
$\sigma_{flux}\ \leq$ 6\%, i.e. the fluxes of each individual spectrum are
on the average accurate better than 6\%. This is in fair agreement with
results of inter-calibrations of spectrophotometric standard stars observed
every night that suggest an accuracy of 5\% or better.

Certain points in Figure~3 are marked with open circles and triangles. The
open circles refer to HM~Sge and V1016~Cyg that have spectra resembling
planetary nebulae. Minimal differences between the $V$ band and eye response
curves around the immensely strong H$\alpha$ and [OIII] lines fully account
for the $\sim$0.7 mag systematic difference of both Asiago and ESO spectra
versus amateurs' estimates. The open triangles refer to V1413~Aql that shows
the same $\sim$0.8 mag systematic difference between the Asiago and ESO
spectra vs. amateurs' estimates: the reason may be an error in the
comparison sequence around V1413~Aql used by the amateurs.

Another independent way to estimate the flux accuracy is offered by the
spectra of the reference objects listed in Table~2. Selecting those not
carrying a variable star name among those observed by Hipparcos/Tycho, we
found
\begin{equation}
V_{T}^{spectra}\ -\ V_{T}^{Tycho}\ = -0.06;\ \ \ \ \sigma\ =\ 0.08
\end{equation}
\begin{equation}
(B-V)_{T}^{spectra} - (B-V)_{T}^{Tycho} = +0.03;\ \ \ \ 
           \sigma\ =\ 0.03
\end{equation}
Here the spectra were convolved with the  $B_T$ and $V_T$ band profiles as
given in the {\sl Hipparcos Catalogue} (ESA SP-1200, June 1997, pag 42;
slightly modified transmission profiles are suggested by Bessell 2000). The
uncertainties in the shutter aperture time ($\sim$0.1 sec in both Asiago and
ESO observations) become notable in the observations of the bright template
stars in Table~2 with an average exposure time of only 2.5 sec. Also,
non-perfect placement of such bright objets on the slit, which is
compensated for faint objects by guiding, may contribute to the scatter in
Eq.(5). Finally, it is worth remembering that Tycho data for bright stars
are accurate to 0.012 mag in $V_T$ and 0.02 mag in $(B-V)_T$.

In light of these considerations it seems safe to argue that the
fluxes in this atlas are generally correct within 
\begin{equation}
\sigma_{flux} \simlt 5\%
\end{equation}
at least for the $B$, $V$ and $R$ band wavelength ranges. A lower
accuracy is probable for faint red objects at $\lambda \leq$3800~\AA\,
or for pure emission line spectra at $\lambda \geq$8700~\AA.

\section{The atlas}

Each spectral observation listed in Tables~1 and 2 is presented by a
separate figure. The format of the figures is identical to Figure~1. UT
dates are in the DD.dd/MM/YY format.

At upper right a full-scale, compressed view of the whole recorded spectrum
is given, and the units for absolute fluxes are indicated. The same units
are valid for the other panels that present zoomed-in portions of the
spectrum.

For ESO spectra the first zoomed-in panel always runs from 3200 to 5250 \AA,
the second from 5150 to 7200 \AA, and the third from 7100 to 9150 \AA. For
Asiago spectra, given the shorter wavelength range covered, two zoomed-in
panels suffice, the first from 3250 to 5600 \AA, the second from 5350 to
7600 \AA.

Stronger emission lines are generally truncated in the zoomed-in panels to
emphasize the visibility of finer details. The full height of emission lines
can be read from the compressed view of the whole spectrum at the upper
right.

\end{document}